\documentclass[acmtosem]{acmtrans2m}

\usepackage{svnkw}
\svnid{$Id: cert-ip.tex 242 2010-07-23 13:52:38Z schallha $}
\usepackage{times}

\usepackage{eurosym}
\usepackage{url}
\usepackage{epsfig}
\usepackage{amssymb}
\usepackage{xspace}
\usepackage{theorem}

\acmVolume{2}
\acmNumber{3}
\acmYear{01}
\acmMonth{09}

\title{Verification Across Intellectual Property Boundaries}

\author{SAGAR CHAKI\\Software Engineering Institute \and
CHRISTIAN SCHALLHART\\Oxford University\and
HELMUT VEITH\\Vienna University of Technology}

\begin{abstract}
  In many industries, the importance of software components provided
  by third-party suppliers is steadily increasing. As the suppliers
  seek to secure their intellectual property (IP) rights, the customer
  usually has no direct access to the suppliers' source code, and is
  able to enforce the use of verification tools only by legal
  requirements. In turn, the supplier has no means to convince the
  customer about successful verification without revealing the source
  code.
  This paper presents an approach to resolve the conflict between the
  IP interests of the supplier and the quality interests of the
  customer. We introduce a protocol in which a dedicated server
  (called the ``amanat'') is controlled by both parties: the customer
  controls the verification task performed by the amanat, while the
  supplier controls the communication channels of the amanat to ensure
  that the amanat does not leak information about the source code.
  We argue that the protocol is both practically useful and
  mathematically sound. As the protocol is based on well-known (and
  relatively lightweight) cryptographic primitives, it allows a
  straightforward implementation on top of existing verification tool
  chains. To substantiate our security claims, we establish the
  correctness of the protocol by cryptographic reduction proofs.
\end{abstract}






\svnid{$Id: macros.tex 241 2010-07-23 13:13:38Z schallha $}

\newcommand{\comment}[1]{}
\newcommand{\m}[1]{\ensuremath{\mathit{#1}}}

\newcommand{\pro}{\ensuremath{\sf Sup}\xspace}
\newcommand{\con}{\ensuremath{\sf Cus}\xspace}
\newcommand{\tbox}{\ensuremath{\sf Ama}\xspace}

\newcommand{\tool}{\ensuremath{{\sf Verifier}}\xspace}
\newcommand{\comp}{\ensuremath{{\sf Compiler}}\xspace}

\newcommand{\progen}{\ensuremath{{\sf Sup\_gen}}\xspace}
\newcommand{\proval}{\ensuremath{{\sf Sup\_val}}\xspace}

\newcommand{\mprogen}{\ensuremath{{\sf MSup\_gen}}\xspace}
\newcommand{\mproval}{\ensuremath{{\sf MSup\_val}}\xspace}

\newcommand{\msup}{\ensuremath{{\sf MSup}}\xspace}

\newcommand{\forge}{\ensuremath{\sf Forge}\xspace}

\newcommand{\result}{\ensuremath{\sf result}\xspace}
\newcommand{\state}{\ensuremath{\sf state}\xspace}

\newcommand{\src}{\ensuremath{\sf source}\xspace}

\newcommand{\bin}{\ensuremath{\sf exec}\xspace}
\newcommand{\cert}{\ensuremath{\sf cert}\xspace}
\newcommand{\ove}{\ensuremath{\sf log}}
\newcommand{\ocon}{{\ensuremath{\ove_{\con}}}\xspace}
\newcommand{\opro}{{\ensuremath{\ove_{\pro}}}\xspace}

\newcommand{\kmpri}{\ensuremath{\m{km}_{\m{priv}}}\xspace}
\newcommand{\kmpub}{\ensuremath{\m{km}_{\m{pub}}}\xspace}

\newcommand{\kcpri}{\ensuremath{\m{ks}_{\m{priv}}}\xspace}
\newcommand{\kcpub}{\ensuremath{\m{ks}_{{pub}}}\xspace}

\newcommand{\krpri}{\ensuremath{\m{kr}_{\m{priv}}}\xspace}
\newcommand{\krpub}{\ensuremath{\m{kr}_{\m{pub}}}\xspace}

\newcommand{\kpri}{\ensuremath{\m{k}_{\m{priv}}}\xspace}
\newcommand{\kpub}{\ensuremath{\m{k}_{\m{pub}}}\xspace}

\newcommand{\round}{\ensuremath{{\sf round}}\xspace}

\newcommand{\sign}{\ensuremath{{\sf csign}}\xspace}
\newcommand{\verifysignature}{\ensuremath{{\sf cverify}}\xspace}
\newcommand{\extract}{\ensuremath{{\sf cextract}}\xspace}

\newcommand{\nop}[1]{}

\newtheorem{definition}{Definition}
\newtheorem{lemma}{Lemma}
\newtheorem{proposition}{Proposition}
\newtheorem{theorem}{Theorem}
\newtheorem{remark}{Remark}
\newtheorem{corollary}{Corollary}
\newtheorem{fact}{Fact}

\begin{document}

\setcounter{page}{1}

\begin{bottomstuff}
  Author's address: chaki@sei.cmu.edu,
  christian.schallhart@comlab.ox.ac.uk, veith@forsyte.at \newline
  A preliminary version of this paper appeared at CAV
  2007~\cite{chaki07:_verif_acros_intel_proper_bound}.\newline
  Supported by the European FP6 project ECRYPT (IST-2002-507932), the
  DFG research grant FORTAS (VE 455/1-1), and the Predictable Assembly
  from Certifiable Components (PACC) initiative at the Software
  Engineering Institute, Pittsburgh, USA.
\end{bottomstuff}
\maketitle

%% FOR NICE URLS IN BIBLIOGRAPHIES
%% Define a new 'leo' style for the package that will use a smaller font.
\makeatletter
\def\url@leostyle{%
  \@ifundefined{selectfont}{\def\UrlFont{\sf}}{\def\UrlFont{\small\ttfamily}}}
\makeatother
%% Now actually use the newly defined style.
\urlstyle{leo}

\svnid{$Id: introduction.tex 246 2010-07-28 15:03:52Z schallha $}
\section{Introduction}
\label{sec:intro}

In classical verification scenarios, the software author and the
verification engineer share a common interest in verifying a piece of
software; the author provides the source code to be analyzed,
whereupon the verification engineer communicates the verification
verdict. Both parties are mutually trusted, i.e., the verification
engineer trusts that she has verified production code, and the author
trusts that the verification engineer will not use the source code for
unintended purposes.

Industrial production of software-intensive technology however often
employs supply chains which render this simple scenario obsolete.
Complex products are being increasingly assembled from multiple
components whose development is outsourced to supplying companies.
Typical examples of outsourced software components comprise embedded
controller software%
% in automobiles and consumer electronics
~\cite{heinecke04:_autom_open_system_archit_indus,broy06:_chall} and
Windows device drivers~\cite{slam-invited}.  Although the suppliers
may use verification techniques for internal use, they are usually not
willing to reveal their source code, as the intellectual property (IP)
contained in the source code is a major asset.

This setting constitutes a principal conflict between the {\em
supplier} \pro who owns the source code, and the {\em customer} \con
who purchases ${\rm o}\;{\rm n}\;{\rm l}\;{\rm y}$ the executable.
While both parties share a basic interest in producing high quality
software, it is in the customer's interest to have the source code
inspected, and in the supplier's interest to protect the source
code. More formally, this amounts to the following basic requirements:

%---------------------------------------------------------------------%
\begin{itemize}
\item[(a)] {\bf Conformance.} The customer must be able to validate
  that the purchased executable was compiled from successfully
  verified source code.

\item[(b)] {\bf Secrecy.} The supplier must be able to ensure that no
  information about the source code besides the verification result is
  revealed to the customer.
\end{itemize}
The main technical contribution of this paper is a new cryptographic
verification protocol tailored for IP-aware verification. Our protocol
is based on standard cryptographic primitives, and satisfies both
requirements with little overhead in the system
configuration. Notably, the proposed scheme applies not only to
automated verification in a model checking style, but also supports a
wide range of validation techniques, both automated and semi-manual.

Our solution centers around the notion of an {\em amanat}. This
terminology is derived from the historic judicial notion of amanats,
i.e., noble prisoners who were kept hostage as part of a contract.
Intuitively, our protocol applies a similar principle: The amanat is
a trusted expert of the customer who settles down in the production
plant of the supplier and executes whatever verification job the
customer has entrusted on him. The supplier accepts this procedure
because (i) all of the amanat's communications are subject to the
censorship of the supplier, and, (ii) 
% the amanat will never return to the customer again.
the amanat will never leave the supplier again.

%---------------------------------------------------------------------%

It is evident that clauses (i) and (ii) make it quite infeasible to
find human amanats; instead, our protocol utilizes a dedicated server
\tbox for this task. The protocol guarantees that \tbox is
simultaneously controlled by both parties: \con controls the
verification task performed by \tbox, while \pro controls the
communication channels of \tbox. To convince \con about conformance,
\tbox produces a cryptographic certificate which proves that the
purchased executable \emph{has been derived by the amanat from the
  same source code as the verification verdict.}

\begin{figure*}[t]
  \center{\fbox{\epsfig{file=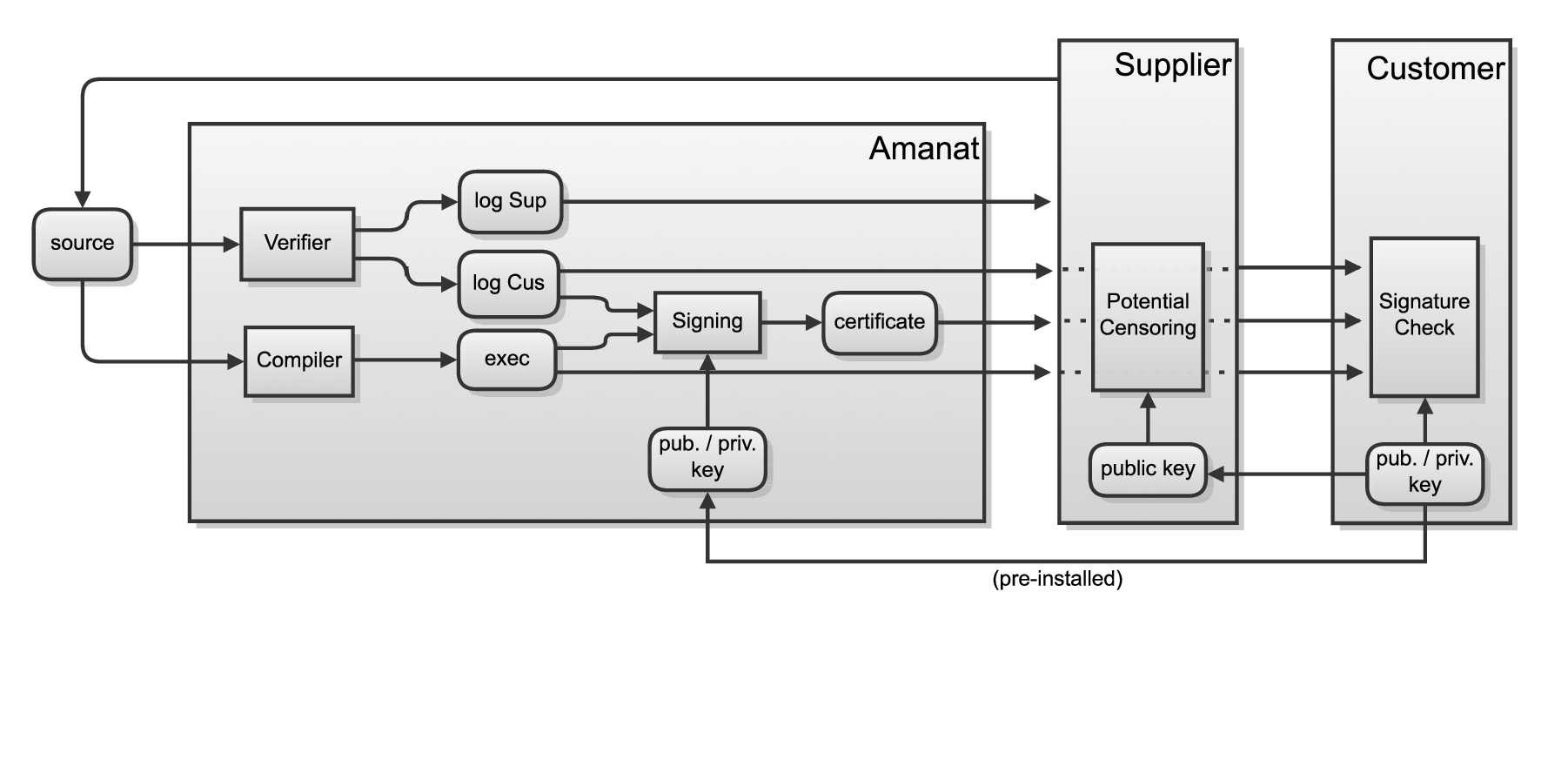,scale=.40}}}
  \caption{A High-Level View of the Amanat
    Protocol\label{fig:overview}}
\end{figure*}

To achieve this goal, we employ public key cryptography; the amanat
uses the secret private key of the customer, and signs outgoing
information with this secret key such that \emph{no additional
  information can be hidden in the signature.}
This enables the supplier to inspect (and possibly block) all outgoing
information, and simultaneously enables the customer to validate that
the certificate indeed stems from the amanat. Hence, the protocol
satisfies clauses (i) and (ii).

Figure~\ref{fig:overview} presents a high-level illustration of the
protocol:
If the code supplier \pro and the customer \con utilize the amanat
protocol, they first install an amanat server \tbox such that (a) \con
is assured that \pro is unable to tamper with the amanat, and (b) \pro
gains complete control over the communication link between \tbox and
\con.
The customer \con equips \tbox with a public/private key pair, such
that \tbox can authenticate its messages to be delivered to
\con. While the public key is handed to \pro, the private key is kept
secret from \pro.
Then, once provided with the tools \comp and \tool, \tbox is ready to
compute certified verification verdicts:
\pro assembles the sources \src and sends them to \tbox. \tbox runs
\tool to obtain the outputs \opro and \ocon, dedicated to \pro and
\con, respectively: \opro may contain IP-critical but
developement-relevant information, while \ocon is only allowed to
contain the verification verdict. Parallel to the verficiation, \tbox
uses \comp to compile the binary \bin.
Using its private key, \tbox computes a certificate \cert which
authenticates \bin together with \ocon as being computed from the same
\src. All these results are returned to \pro.
Then \pro checks whether all computations have been performed as
expected, i.e., whether \bin and \ocon resulted from respective
invocations of \comp and \tool, and whether \cert has been computed
properly.
Since the computation of \cert involves random bits, the amanat
protocol allows \pro to check that \tbox chose these random bits
\emph{before} accessing \src---such that they cannot contain any
information on \src.
If these checks succeed, \pro is assured that the secrecy of its IP
has been preserved.
After evaluating \opro and \ocon regarding its content, \pro decides
whether to forward \bin, \ocon, and \cert to \con across the IP
boundary or not.
If the results are forwarded to \con, \cert is checked by \con and if
the certificate is valid, \con accepts \bin and \ocon as being
conformant.

%% To achieve this goal, we use public key cryptography; the amanat
%% uses the secret private key of the customer, and encrypts outgoing
%% information with this secret key. This enables the supplier to {\em
%% decrypt} (and possibly block) all outgoing information, and
%% simultaneously enables the customer to validate that the certificate
%% indeed stems from the amanat. Thus, the amanat protocol achieves the
%% two requirements above. Figure~\ref{fig:overview} presents a
%% high-level illustration of the protocol.

%%%%%%%%%%%%%%%%%%%%%%%%%%%%%%%%%%%%%%%%%%%%%%%%%%%%%%%%%%%%%%%%%%%%%%%%%%%%%%%%

\subsection{Verification by Model Checking and Beyond}

Motivated by discussions with industrial collaborators, we primarily
intended our protocol to facilitate software model checking across IP
boundaries in a B2B setting where suppliers and customers are
businesses. Our guiding examples for this setting have been Windows
device drivers and automotive controller software, for which our
protocols are practically feasible with state-of-the-art technology.

Software model checking is now able to verify important properties of
simply structured code
\cite{slam01,blast02,icse03,DBLP:conf/tacas/ClarkeKL04,DBLP:conf/padl/PodelskiR07}.
Most notably, SLAM/SDV is a fully automatic tool for a narrow
application area, and we expect to see more such tools.  Note that SDV
has built-in specifications because the device drivers access and
implement a clearly defined API. Other tools such as
Terminator~\cite{Terminator} and Slayer~\cite{Slayer} do not require
specifications as they are built to verify specific critical
properties -- termination and memory-safety, respectively.

Software engineering has become essential in the automotive
industries:
For example, the current BMW 7 series implements about 270 user
observable features deployed on 67 embedded platforms with
approximately 65 MB of binary
code~\cite{pretschner07:_softw_engin_autom_system}.
In 2002, an estimated value of \euro{} 127 billion was created by
electric, electronics, and software components within the automotive
domain---and by 2015, this amount is expected to rise up to \euro{}
316 billion~\cite{dannenberg04:_comin_age_collab_autom_indus}. 
%
%% Within this segment of automotive value creation, software is
%% expected to make up approximately 40\% by
%% 2010~\cite{hardung04:_reuse_softw_distr_embed_autom_system}.
%
As evolution and integration have been identified as predominant
issues in automotive system
software~\cite{pretschner07:_softw_engin_autom_system},
a number of initiatives have been started to establish standardized
and industry-wide accepted computing environments, most notably the
Automotive Open System Architecture (AUTOSAR)~\cite{07:_autos_consor},
the Open Systems and the Corresponding Interfaces for Automotive
Electronics (OSEK)\footnote{The acronym is derived from the German
  project name ``Offene Systeme und deren Schnittstellen f\"ur die
  Elektronik in
  Kraftfahrzeugen''.}~\cite{07:_offen_system_schnit_elekt_kraft}, and
the Japan Automotive Software Platform Architecture
(JASPAR)~\cite{07:_japan_autom_softw_platf_archit_jaspar}.
These standardization efforts originate in the need to \emph{integrate
  software components developed by various companies along a deep
  supply chain.}
This task demands for verification of both, general requirements to be
satisfied by every component and specific requirements dedicated to
individual components---in presence of heavy IP interests which must
be respected and
protected~\cite{pretschner07:_softw_engin_autom_system}.

%% Automotive software is similar to device drivers in that it also
%% accesses standardized APIs~\cite{osek,07:_autos_consor,jaspar}.

The amanat protocol provides a general framework to perform source
dependent, validation tasks in presence of IP boundaries and is
therefore \emph{not restricted} to pure verification:
%% 
%% 
%% But the amanat protocol is \emph{not restricted} to pure verification
%% tasks: 
%
For example, it may be necessary for customers and suppliers to
communicate some software design details without revealing the
underlying source code. In this case, the supplier can decide to
reveal a blueprint of the software, and the amanat can certify the
accuracy of the blueprint by a mutually agreed algorithm.
This is possible, because the amanat can run any
verification/validation tool whose output does not compromise the
secrecy of the source code. For example, the amanat protocol is
applicable to the following techniques:

%---------------------------------------------------------------------%
\begin{enumerate}
\item apply static analysis tools such as 
ASTREE~\cite{astree} and
  TVLA~\cite{tvla}.

\item \label{enum:theorem-prover} check the correctness of a proofs
  provided by \pro, given in, e.g., PVS, ISABELLE, Coq or another
  prover~\cite{provers}.

\item evaluate worst case execution times experimentally~\cite{kirner}
  or statically~\cite{wilhelm}.

\item generate white box test
  cases~\cite{holzer08,holzer10:_how_did_you_specif_your_test_suite},
  and execute them.

\item validate that the source code comes with a test suite which
  satisfies previously agreed coverage criteria, using
  e.g.~CoverageMeter~\cite{CoverageMeter} or
  BullseyeCoverage~\cite{BullseyeCoverage}.

\item check that the source code is syntactically safe, e.g.~using
  LINT~\cite{60274}.

\item \label{enum:code-metrics} compute numerical quality and quantity
  measures which are agreed between \pro and \con, e.g. nesting depth,
  LOC, etc.

\item compare two versions of the source code, and quantify the
  difference between them; this is important in situations where \pro
  claims charges for a reimplementation.

\item check if third party IP is included in the source code,
  e.g. libraries etc.

\item ensure that certain algorithms are (not) used.\label{item:3}

\item check that the source code is well-documented.\label{item:4}

\item validate the development steps by analyzing the CVS or SVN tree.

\item ensure compatibility of the source code with language standards.
\end{enumerate}

We note that in all these scenarios the code supplier {\em bears the
  burden of proof}: \tbox is provided with \src and must be able to
verify---using the tool \tool alone---that \src and the resulting
binary \bin are indeed satisfying the specified requirements.
To this end, \src may also contain auxiliary information supporting
\tool in computing the verification verdict. For example, if \tool
uses a theorem prover (as in item~\ref{enum:theorem-prover} above),
then \src contains additionally a proof which actually shows the
conformance of the source code to the specification.
For most cases mentioned above, at least some auxiliary information is
provided in \src, such as command line options, abstraction functions,
or test cases.

It is essential that \tbox (more precisely, the agreed upon tool
\tool) only accepts auxiliary information which helps in proving the
required correctness properties more efficiently but does not take
influence on the resulting verdict.
For example, a proof for a theorem prover will help \tbox to verify
the required properties---but if the provided proof is wrong or
misleading then \tbox will not produce a wrongly positive verdict.

\subsection{ Security of the Amanat Protocol} 

We present in Sections~\ref{sec:protocol-secrecy} and
\ref{sec:protocol-conformance} cryptographic proofs for the {\em
  secrecy} and {\em conformance} of the amanat verification
protocol. Stronger than term-based proofs in the Dolev-Yao
model~\cite{dolev81:_secur_of_public_key_protoc}, these proofs assure
that under standard cryptographic assumptions, randomized polynomial
time attacks against the protocol (which may involve e.g.~guessing the
private keys) can succeed only with negligible probability (for a
technical definition of negligible probability, see the discussion
following Definition~\ref{def:semantic-security}).
The practical security of the protocol is also ensured by its
simplicity: As the protocol is based on well-known cryptographic
encryption and signing schemes, it can be  implemented with reasonable
effort.

The cryptographic protocols require \con and \tbox to \emph{share a
  secret} unknown to \pro, namely the private key of the customer;
this secret enables \tbox to authenticate its verdict to \con and by
computing the certificate. Consequently, the cryptographic proofs need
to assume a physical system configuration and infrastructure where
\tbox can neither be reverse-engineered, nor closely monitored by the
supplier. On the other hand, from the point of view of \pro, \tbox is
an untrusted black box with input and output channels. For secrecy,
the supplier requires ownership of \tbox to make sure it will not
return to the customer after verification. There are two natural
scenarios to realize this hardware configuration:

\begin{itemize}
\item[A] \tbox is physically located at the site of a trusted third
  party. All communication channels of \tbox are hardwired to go
  through a second server, which acts as communication filter of the
  supplier, cf.~Figure~\ref{fig:overview}.
\end{itemize}

\noindent While scenario A involves a trusted third party, its role is
limited to providing physical security for the servers and requires no
expertise beyond server hosting. For the supplier, scenario A has the
disadvantage that the encrypted source has to be sent to the third
party, and thus, to leave the supplier site.

\begin{itemize}
\item[B] \tbox is physically located at the site of the supplier, but
  in a sealed location or box whose integrity is assured through,
  e.g., regular checks by the customer, a third party, a traditional
  alarm system, or the use of sealed
  hardware~\cite{DBLP:conf/vlsid/RaviRC04}. All communication channels
  of \tbox are hardwired to the communication filter of the supplier.
\end{itemize}

\noindent 
We believe that in our B2B settings, scenario B is practically
feasible: It only requires that the seal is checked after verification
and before deployment. Thus, there is no business incentive for the
supplier to break the seal.

The supplier has total control over the information leaving the
production site. Thus, it can also prevent attempts by the amanat to
send messages at specific time points and to thereby leak information.
\emph{The supplier can read, delay, drop, and modify all outgoing
  messages}---which is a convincing and non-technical argument that no
sensitive information is leaking. In our opinion, this simplicity of
the amanat protocol is a major advantage for practical application.

\medskip

\paragraph*{Organization of the Paper}
In Section~\ref{sec:related}, we survey related work and discuss
alternative approaches to the amanat protocol. Afterwards in
Section~\ref{sec:proto}, we introduce the relevant tools and
cryptographic primitives, followed by a brief protocol overview, and a
detailed description of the protocol.
The secrecy and conformance of the protocol are shown in
Sections~\ref{sec:protocol-secrecy}
and~\ref{sec:protocol-conformance}, respectively, and the paper is
concluded with Section~\ref{sec:conclusion}.

While the proof on secrecy follows an intuitive argument, the proof of
conformance is technically more involved. We therefore start
Section~\ref{sec:protocol-conformance} with the theorem stating the
conformance of the protocol, even before precisely defining the
underlying cryptographic assumptions. Subsequently, we introduce these
assumptions in Section~\ref{sec:security-properties}, give an overview
on the proof in Section~\ref{sec:proof-overview} and present its
details in Sections~\ref{sec:modeling-supplier}
to~\ref{sec:proof-theorem-ref}.

%%% Local Variables:
%%% mode: latex
%%% TeX-master: "../cert-ip"
%%% ispell-local-dictionary: "american"  ***
%%% End:

% LocalWords:  amanat

\svnid{$Id: related-work.tex 241 2010-07-23 13:13:38Z schallha $}
\section{Related Work and Alternative Solutions}
\label{sec:related}

The last years have seen renewed activity in the analysis of
executables from the verification and programming languages community.
Despite remarkable advances (see e.g.
~\cite{reps-vmcai,debray,reps-aplas,cifuentes,veithkinder-fmcad2010}),
the computer-aided analysis of executables remains a hard problem;
natural applications are reverse engineering, automatic detection of
low level errors such as memory violations, and malicious code
detection~\cite{somesh-analysis,kinder08:_detec_malic_code_model_check}.
The technical difficulties in the direct analysis of executables are
often exacerbated by code obfuscation to prevent reverse engineering,
or, in the case of malware, recognition of the malicious code.
Although dynamic analysis~\cite{ColinM05} and 
approaches dealing with black box
systems~\cite{LeeYannakakis94,LeYa:survey,peled-black-box} are
relatively immune to obfuscation, they are limited either in the range
of systems they can deal with or in the correctness properties they
can assure.
%% 
%% black box testing~\cite{LeeYannakakis94,LeYa:survey} are relatively
%% immune to obfuscation, they only give a limited assurance of system
%% correctness.
%% 
%% and black box
%% checking\cite{peled-black-box} 

The current paper is orthogonal to executable analysis. We consider a
scenario where the software author is willing to assert the quality of
the source code by formal methods, but does not or provide the source
code to the customer. 
%
%% It is evident that the visibility of the source code to the amanat
%% and the cooperation of the software author/supplier significantly
%% increase the leverage of formal methods.

While proof-Carrying Code~\cite{necula97proofcarrying} is able to
generate certificates for binaries, it is only applicable for a
restricted class of safety policies. More importantly, a proof for
non-trivial system properties will explain---for all practical
purposes---the internal logic of the binary, and thus, publishing this
proof is tantamount to losing intellectual property.

%On the computer security side, we have designed the protocol to
%employ established classical concepts, in particular an asymmetric
%encryption and signature scheme
%  \cite{cramer00:_signat_schem_based_strin_rsa_assum} based upon
%  RSA~\cite{rivest78:_method_obtain_digit_signat_public_key_crypt} and
%  SHA~\cite{standards95:_nist_fips_pub_secur_hash_stand}.
 % 
The current paper takes an engineer's view on computer security, as it
exploits the conceptual difference between source code and executable.
% 
%% The current paper takes an engineer's view on computer security.
%% The results of the paper are quite specific to verification, as it
%% exploits the conceptual difference between the source code and the
%% executable.
%
While we are aware of advanced methods such as secure multiparty
computation~\cite{secure-multi-party} and zero-knowledge
proofs~\cite{704888}, we believe that they are impracticable for our
problem. To implement secure multiparty computation, it would be
necessary to convert significant parts of the model checking tool
chain into a Boolean circuit which is not a realistic option.
To apply zero-knowledge proofs, one would require the verification
tools to produce highly structured and detailed formal proofs. Except
for the provers in item 2 of the list in Section \ref{sec:intro}, it
is impractical to obtain such proofs by state of the art technology.
Finally, we believe that any advanced method without an intuitive
proof for its secrecy will be heavily opposed by the supplier---and
might therefore be hard to establish in practice.
Thus, we are convinced that the conceptual simplicity of our protocol
is an asset for practical applicability.

\nop{
 In our understanding, current industrial practice is based on
semi-legal reverse engineering

Common techniques for verification across IP boundaries attain the
requirements of the supplier and the consumer only in a very limited
way.

On the one hand, as a technical measure, the consumer can apply
verification techniques such as black box testing which do not
require knowledge of the source code. More precise techniques, such
as model checking, however, are known to be significantly harder for
executables than for source code so that only few properties can be
checked realistically.

On the other hand, a legal contract may require the supplier to
employ certain verification techniques or software quality
standards. This traditional approach is useful to settle liability
issues {\em after} the fact, e.g., when a product has experienced
severe problems. In this case, a court appointed expert will
eventually check if the supplier has fulfilled the contract, but the
commercial loss incurred by the consumer up to this time point may
be intolerably large.

If the contract instead requires frequent manual inspections of the
source code, then the parties need to hire an expert as a trusted
third party on a regular basis.  As the suppliers have to reveal
their source code to this expert and are subject to constant
supervision, they usually do not favor this approach. }

%%% Local Variables:
%%% mode: latex
%%% TeX-master: "../cert-ip"
%%% ispell-local-dictionary: "american"  ***
%%% End:

\svnid{$Id: protocol.tex 241 2010-07-23 13:13:38Z schallha $}
\section{The Amanat Protocol}
\label{sec:proto}

The amanat protocol aims to resolve the conflict between the customer
\con who wants to verify the source code, and the supplier \pro who
needs to protect its IP. To this end, the amanat \tbox computes a
certificate which contains a verdict on the program correctness, but
does not reveal any information beyond the verdict itself.

\subsection{Requirements and Tool Landscape}

To make the protocol requirements more precise, we fix some notation
and assumptions about the tool landscape. We restrict our tools to run
in deterministic polynomial time to ensure that their computations can
be efficiently reproduced. 
When we deal with higher runtime complexities, we pad \src, i.e., add
a string long enough to upper-bound the runtime of all tools with a
polynomial in the padded length of \src. By adding random seeds to
\src, we can also integrate randomized tools into our framework.

\begin{definition}[Compiler]
  The {\em compiler} \comp translates an input \src into an executable
  $\bin = \comp(\src)$ in deterministic polynomial time.
\end{definition}
Note that \comp does not take any further input. In practice, this
means that \src is a directory tree and that \comp is a tool chain
composed of compiler, linker etc.

\begin{definition}[Verification Tool]
  The {\em verification tool} \tool takes the input \src and computes
  in deterministic polynomial time two verification verdicts, \opro
  and \ocon, i.e., $\left<\opro,\ocon\right>=\tool(\src)$.
\end{definition}
Here, \opro is a detailed verdict for the supplier possibly containing
IP-critical information such as counterexamples or witnesses for
certain properties. The second output \ocon in contrast contains only
uncritical verification verdicts which \pro and \con have agreed upon
beforehand.

Similar as for the compiler, we assume that \tool does not take any
inputs besides \src. Thus, the specification is included into
\src---allowing \tbox to output the verification results together with
their specifications into \ocon. Hence, \pro can check which
properties have been verified by \tbox. All auxiliary information
necessary to run \tool is provided by \pro as part of \src, such as
command line parameters or abstraction functions.

As in the case of \comp, \tool is not restricted to consist of a
single tool. On the contrary, \tool can comprise a whole set of
verification tools, as long as they are tied together to produce a
single pair $\left<\opro,\ocon\right>$ of combined outputs.

Having fixed environment and notation, we paraphrase the requirements
in a more precise manner:

%---------------------------------------------------------------------%
\begin{definition}[Conformance]
  \label{def:conformance}
  An execution of the amanat protocol is \emph{conformant,} if the
  delivered binary \bin and verdict \ocon have been produced from the
  same \src.
\end{definition}

\begin{definition}[Secrecy]
  \label{def:secrecy}
  An execution of the amanat protocol ensures {\em secrecy}, if all
  information provided to \con in the course of the protocol is either
  directly contained in or implied by \bin and \ocon.
\end{definition}

The goal of the Amanat protocol is to give mathematical guarantees for
these two properties for all (but a negligible fraction) of protocol
executions---based on two assumptions: First, the communication
channels between \pro, \con, and \tbox must be secure, i.e., the
protocol is not designed to cope with orthogonal risks such as
eavesdropping or malicious manipulations on these channels.
Second, we assume that all ingoing and outgoing information for \tbox
is controlled by \pro, i.e., \pro can manipulate all data exchanged
between \tbox and \con.

%% \begin{remark}[Security of the Communication Channels]
%%   \label{rem:security-of-the-communication-channels}
%%   We assume that the \emph{communication channels} between \pro, \con,
%%   and \tbox are secure, i.e., the protocol is not concerned with
%%   eavesdropping or malicious manipulations on these channels.
%%   %
%%   Moreover, all ingoing and outgoing information for \tbox is
%%   controlled by \pro, i.e., \pro can manipulate all data exchanged
%%   between \tbox and \con.
%% \end{remark}

We note that some of the possible verification tasks discussed in
Section~\ref{sec:intro}---in particular~\ref{enum:code-metrics},
\ref{item:3}, \ref{item:4}---are concerned with non-functional
properties of the source code which do not affect the executable
produced by the compiler. However, the conformance property
\emph{only} proves to the customer that the obtained binary and its
verification verdict stem from the same source code.
Thus, in the case of a legal conflict, a court can \emph{only} require
the supplier to provide a source code which (i) compiles into the
purchased executable, and (ii) produces the same verification output
\ocon. There is no mathematical guarantee however, that the revealed
code will be {\em identical} to the original code.
For example, if the verdict only indicates whether the delivered
executable is deadlock free or not, then the supplier can decide to
not reveal its original source but to provide an obfuscated source
which does not contain any comments and has all identifiers renamed
into random strings.

\begin{remark}[Preventing Obfuscation]
  In situations where it is necessary to obtain the original source,
  we equip \tool with (i) a check that \src does not have an
  obfuscated appearance, and (ii) a secure hash computation which
  appends the hash of \src to \ocon---leaving the protocol itself
  unchanged.
  % 
  %% To force \pro to disclose its original source in case of a legal
  %% conflict, we let \tool check that \src does not have an obfuscated
  %% appearance, and include a secure hash of \src in \ocon. 
  % 
  Then, in case of a conflict, \pro must provide its original source
  to match the secure hash.
\end{remark}

\subsection{Cryptographic Primitives}
\label{sec:crypt-prim}

Before we formally describe the primitives for encrypting, decrypting,
signing and verifying messages, we note that the underlying algorithms
are not deterministic but \emph{randomized.}  This randomization is a
countermeasure to attacks against naive implementations of RSA and
other schemes which exploit algebraically related messages, see for
example~\cite{dolev00:_non_malleab_crypt}. In many protocols, the
randomization can be treated as technical detail, as each participant
can locally generate random values. But in our protocol, we must
ensure that the signatures generated by \tbox do not contain hidden
information for \con---and must hence deal with randomization
explicitly: Using methods from
steganography~\cite{petticolas00:_infor_hidin_techn_stegan_digit_water},
\tbox could encode source code properties into allegedly randomly
generated bits. To preclude this possibility, our protocol forces
\tbox to commit its random bits {\em before} it sees the source code.

Below, we define schemes for encrypting and signing messages. In case
of the signature scheme, we also add procedures with explicit
randomization parameters.
As both schemes use an asymmetric key-pair, we assume that the same
pair $\langle \kpri, \kpub \rangle$ can be used for both. This can
be easily achieved by combining the key-pairs for both schemes into a
single pair.

\begin{definition}[Public-Key Encryption Scheme]
  \label{def:encryption}
  Given a \emph{key pair} $\langle \kpri, \kpub \rangle$, we
  define the encryption and decryption and their respectively required
  computational complexity bound (with respect to length of their
  inputs and security parameters) as follows:
  \begin{itemize}
  \item \textbf{Encryption:} For a \emph{plaintext} message $m$, we
    write $c=\kpub(m)$ to denote the \emph{encryption} of $m$ with
    key $\kpub$ yielding the \emph{ciphertext} $c$
    (probabilistic polynomial time).

  \item \textbf{Decryption:} Similarly, $m=\kpri(c)$ denotes the
    \emph{decryption} of the ciphertext $c$ with key $\kpri$
    resulting again in the original message $m$
    (deterministic polynomial time).
  \end{itemize}
\end{definition}

\begin{definition}[Public-Key Signature Scheme]
  \label{def:signatures}
  Given a \emph{key pair} $\langle \kpri, \kpub \rangle$, we
  define the following operations running in deterministic polynomial
  time (with respect to length of their inputs and the security
  parameter):
  \begin{itemize}
  \item \textbf{Signature Generation:} We write $s=\sign(\kpri,m,R)$
    for the \emph{signature} $s$ of a message $m$ signed with key
    $\kpri$ and generated with random seed $R$.

  \item \textbf{Signature Verification:}
    $\verifysignature(\kpub,m,s)$ denotes the verification result of
    a signature $s$ for message $m$ with key $\kpri$. The
    verification succeeds, iff there exists a random seed $R$ such
    that $s=\sign(\kpri,m,R)$ holds.

  \item \textbf{Random Seed Extraction:} We write $R=\extract(s)$ for
    $s=\sign(\kpri,m,R)$ to extract the random seed $R$ used in a
    signature $s$ generated for message $m$ with key $\kpri$.

  \item \textbf{Signature Verification with Fixed Random Seed:} 
    We write $\verifysignature(\kpub,m,s,R)$ to check whether a 
    signature $s$ for message $m$ and $\kpri$ has been generated
    with random seed $R$, i.e., $\verifysignature(\kpub,m,s,R)$
    succeeds iff $s=\sign(\kpri,m,R)$ holds.
  \end{itemize}
\end{definition}

Thus, besides standard signature generation and verification with
$\sign(\kpri,m,R)$ and $\verifysignature(\kpub,m,s)$,
respectively, we require the existence of two additional procedures:
The first one, $\extract(s)$ extracts the random seed $R$ used in
signature $s$, and the second one, $\verifysignature(\kpub,m,s,R)$
verifies that signature $s$ has been generated with random seed $R$.

Aside providing these interfaces, suitable cryptographic primitives
must also satisfy the relevant security properties, as defined in
Section~\ref{sec:security-properties}:
In case of the encryption scheme, our requirements are fairly standard
and are satisfied by a number of encryption schemes, e.g.~one can use
ElGamal encryption~\cite{elgamal85:_public_key_crypt_signat_schem}.
For the signature scheme, we propose to use
\cite{cramer00:_signat_schem_based_strin_rsa_assum} which is based
upon RSA~\cite{rivest78:_method_obtain_digit_signat_public_key_crypt}
and SHA~\cite{standards95:_nist_fips_pub_secur_hash_stand} and allows
to implement all operations described above.

\subsection{Summary Description of the Protocol}

Our protocol is based on the principle that \con\ trusts \tbox, and
thus, \con believes that a verification verdict \ocon originating from
\tbox is \emph{conformant} with a corresponding binary \bin.
Therefore, \con and \pro install \tbox at \pro's site such that \pro can
use \tbox to generate trusted verification verdicts subsequently.
At the same time, \pro controls all the communication to and from
\tbox and consequently \pro is able to prohibit the communication of
any piece of information beyond the verification verdict, i.e., \pro
can enforce the \emph{secrecy} of its IP.
To ensure that \pro does not alter the verdict of \tbox, \tbox signs
the verdicts with a key which is only known to \tbox and \con but not
to \pro. Also, to ensure that the tools \comp and \tool given to \tbox
are untampered, \pro must provide certificates which guarantee that
these tools have been approved by \con.

A protocol based on this simple idea does ensure the conformance
property, but a naive implementation with common cryptographic
primitives fails to guarantee secrecy:
As argued above, the certificates generated by \tbox involve {random
  seeds}, and \pro \emph{cannot check} these random seeds for hidden
information.
%
%On the contrary, \tbox could extract some critical information from
%the \src to be validated and encrypt this information to obtain an
%allegedly random number. \pro could not find the hidden piece of
%information and would forward the certificates to \con. Then, \con
%would extract the involved random seeds and decrypt the seemingly
%random number in order to recover the hidden information on the
%\src.
%
In our protocol, to prohibit such hidden transmission of
information, \tbox is not allowed to generate the required random
seeds after it has accessed \src.
Instead, \tbox generates a large supply of random seeds {\em before}
it has access to \src, and sends them to \pro. In this way, \tbox
commits to the random seeds. Later, \pro checks that \tbox used
exactly these random values. Thus, \tbox is not able to encode any
information about \src into these seeds.

The only remaining problem is that \pro is \emph{not allowed to know
  the random seeds in advance,} since it could use this knowledge to
compromise the cryptographic security of the certificates computed by
\tbox. Therefore, \tbox encrypts each random seed with a specific key
before transmitting them to \pro, and reveals the corresponding key
when it uses one of its seeds.

\subsection{Detailed Protocol Description}

Our protocol consists of three phases, namely the \emph{installation,}
\emph{session initialization,} and \emph{certification.}

\textbf{Installation Phase:} \con initializes \tbox with a master key
pair $\langle \kmpri,\kmpub \rangle$ which is used later to exchange a
session key pair. Then, \tbox is transported to and installed at the
designated site. All further communication between \tbox and \con is
controlled by \pro.

\medskip
\begin{enumerate}\itemsep=.4em
\item[\bf I1] {\em\bfseries Master Key Generation} \hfill [ \con]\\
  \con generates the master keys $\langle \kmpri,\kmpub \rangle$
  and initializes \tbox with them.
\item[\bf I2] {\em\bfseries Installation of the Amanat} \hfill [ \pro, \con]\\
  \tbox is installed at \pro's site and \pro receives $\kmpub$.
\end{enumerate}
\medskip

\textbf{Session Initialization Phase:}
After installation, \pro and \con must agree on a specific \tool and
\comp. Once \tool and \comp have been fixed, the session
initialization phase starts: First, \con generates a new pair of
session keys $\langle \kcpri,\kcpub \rangle$ and sends them to \tbox
via \pro---having encrypted the private key \kcpri beforehand. Then,
the new session keys are used to produce certificates $\cert_{\tool}$
and $\cert_{\comp}$ for \tool and \comp, respectively. \pro checks the
contents of the certificates. If they are valid, it uses them to setup
\tbox with \tool and \comp.
\tbox in turn accepts \tool and
\comp if their certificates are valid.

In the last initialization step, \tbox generates a supply of random
seeds $R_1,\dots,R_t$ for $t$ subsequent executions of the
certification phase. It also generates a sequence of key pairs
$\langle \krpri^1,\krpub^1\rangle,\dots,\langle
\krpri^t,\krpub^t\rangle$ for each random seed $R_i$. \tbox finally
encrypts each random seed to obtain and send $\krpub^i(R_i)$ to \pro.
\tbox and \pro both maintain a variable \round which is initialized to
0 and incremented by 1 for each execution of the certification phase.

%---------------------------------------------------------------------%
\medskip
\begin{enumerate}\itemsep=.4em
\item[\bf S1] {\em\bfseries Session Key Generation} \hfill [ \con, \pro]\\
\con generates the session keys $\langle \kcpri,\kcpub \rangle$
  and sends $\kmpub(\kcpri)$ and \kcpub to \pro. \pro forwards
  $\kmpub(\kcpri)$ and \kcpub unchanged to \tbox.

\item[\bf S2] {\em\bfseries Generation of the Tool Certificates} \hfill [ \con]\\
\con computes the certificates
  \begin{itemize}
  \item $\cert_{\tool}=\sign(\kcpri,\tool)$ and
  \item $\cert_{\comp}=\sign(\kcpri,\comp)$.
  \end{itemize}
  \con sends both certificates to \pro.

\item[\bf S3] {\em\bfseries Supplier Validation of the Tool Certificates} \hfill [ \pro]\\
\pro checks the contents of the certificates, i.e., \pro checks
  that
  \begin{itemize}
  \item $\verifysignature(\kcpub,\tool,\cert_{\tool})$ and
  \item $\verifysignature(\kcpub,\comp,\cert_{\comp})$ succeed.
  \end{itemize}
  If one of the checks fails, \pro aborts the protocol.

\item[\bf S4] {\em\bfseries Amanat Tool Transmission} \hfill [ \pro]\\
  \pro sends \tool, \comp, and the certificates $\cert_{\tool}$ and
  $\cert_{\comp}$ to \tbox.

\item[\bf S5] {\em\bfseries Amanat Validation of the Tool Certificates} \hfill [ \tbox]\\
\tbox checks whether \tool and \comp are properly certified,
  i.e., it checks whether
  \begin{itemize}
  \item $\verifysignature(\kcpub,\tool,\cert_{\tool})$ and
  \item $\verifysignature(\kcpub,\comp,\cert_{\comp})$ succeed.
  \end{itemize}
  If this check fails, \tbox refuses to process any further input.

\item[\bf S6] {\em\bfseries Amanat Random Seed Generation} \hfill [ \tbox]\\
  \tbox generates
  \begin{itemize}
  \item a series of random seeds $R_1,\dots,R_t$ together with a
    series of corresponding key pairs $\langle \krpri^1,\krpub^1
    \rangle,\dots, \langle \krpri^t,\krpub^t \rangle$, 
  \item encrypts the random seeds with the corresponding keys
    $\krpub^i(R_i)$ for $i=1,\dots,t$, and
  \item initializes round counter $\round=0$.
  \end{itemize}
  \tbox sends $\krpub^i(R_i)$ and $\krpub^i$ for $i=1,\dots,t$ to
  \pro.
\end{enumerate}
\medskip

\textbf{Certification Phase:}
\tbox is now ready for the certification phase, i.e., it will accept
\src to produce a certified verdict on \src which can be forwarded to
\con and whose trustworthy origin can be checked by \con.

During certification, \tbox runs \tool and \comp on \src and generates
a certificate \cert for the binary \bin and the verification output
\ocon dedicated to \con.
The certificate is based upon the random seed $R_\round$ which \tbox
committed to use in this round during session initialization.
\tbox sends the certificate \cert, the outputs \opro and \ocon, and
the key $\krpri^\round$ to \pro.

To validate secrecy, \pro computes the random seed
$R_\round=\krpri^\round(\krpub(R_\round))$ which \tbox supposedly used
for the generation of \cert.
Then \pro checks that the certificate \cert is valid and based upon
the random seed $R_\round$.
If this is the case, \tbox cannot hide any unintended information in
the certificates. Otherwise, if the checks fails, \pro aborts the
protocol.
If \pro proceeds and the obtained verdict is good enough, it forwards
the results to \con. Otherwise, it might revise \src and start a new
certification phase. Finally, when \con receives \bin, \ocon, and
\cert, it checks conformance \bin and \ocon using \cert.

%---------------------------------------------------------------------%

\medskip
\begin{enumerate}\itemsep=.4em
\item[\bf C1] 
  {\em\bfseries Source Code  Transmission} \hfill [ \pro]\\
  \pro sends \src to \tbox.

\item[\bf C2] {\em\bfseries Source Code Verification by the Amanat} \hfill [ \tbox]\\
  \tbox computes
  \begin{itemize}
  % \item the private key $\krpri=\kcpri(\kcpub(\krpri))$,
  %\item the random seeds $R_1=\krpri(\krpub(R_1))$ and
  %  $R_2=\krpri(\krpub(R_2))$,
  \item the verdict $\left\langle \opro,\ocon\right\rangle=\tool(\src)$ of \tool on
    \src,
  \item the binary $\bin=\comp(\src)$, 
  \item increments the round counter \round, and 
  \item $\cert=\sign(\kcpri,\left\langle\bin,\ocon\right\rangle,
    R_\round)$.
  \end{itemize}
  \tbox sends \bin, \opro, \ocon, \cert, and $\krpri^\round$ to \pro.

\item[\bf C3] 
  {\em\bfseries Secrecy Validation} \hfill [ \pro]\\
  Upon receiving \bin, \opro, \ocon, \cert, and $\krpri^\round$, \pro
  \begin{itemize}
  \item decrypts the random seed with
    $R_\round=\krpri^\round(\krpub^\round(R_\round))$,
  \item checks whether $\bin=\comp(\src)$ and
    $\left<\opro,\ocon\right>=\tool(\src)$ hold, and
  \item verifies that
    $\verifysignature(\kcpub,\left\langle\bin,\ocon\right\rangle,\cert,R_\round)$ succeeds.
  \end{itemize}
  If the checks fails, \pro\ {\bf concludes that the secrecy
    requirement has been violated}, and refuses to proceed with the
  protocol.

  Otherwise, \pro evaluates \ocon and \opro and decides whether to
  deliver \bin, \ocon, and \cert to \con in step {\bf C4} or whether
  to abort the protocol.

\item[\bf C4] 
  {\em\bfseries Conformance Validation} \hfill [ \con]\\
  Having received \bin, \ocon, and \cert, \con verifies that
  certificate is valid, i.e., that
  $\verifysignature(\kcpub,\left\langle\bin,\ocon\right\rangle,\cert)$
  succeeds.
  
  If the checks fails, \con\ {\bf concludes that the conformance
    requirement has been violated}, and refuses to proceed with the
  protocol.

  Otherwise \con evaluates the contents of \ocon and decides whether
  the verification verdict supports the purchase of the product \bin.
\end{enumerate}
\medskip

%%% Local Variables:
%%% mode: latex
%%% TeX-master: "../cert-ip"
%%% ispell-local-dictionary: "american"  ***
%%% End:

% LocalWords:  amanat

\svnid{$Id: protocol-secrecy.tex 241 2010-07-23 13:13:38Z schallha $}
\section{Protocol Secrecy}
\label{sec:protocol-secrecy}

We designed the amanat protocol aiming at a simple and intuitive
argument for its secrecy, as such a straightforward proof is a
prerequisite to convince a code supplying company that the protocol
keeps their highly valued IP assets safe.
For the same reason, we do not rely on any cryptographic assumptions
to prove the secrecy of the amanat protocol. 

The idea behind this proof is straightforward: As the certificate
\cert is the only place to transmit additional information from \tbox
to \con, we make sure that \cert can be computed without \emph{knowing
  the source itself.} Hence, no information on the source can be
possibly hidden in \cert.

\begin{theorem}[Secrecy]
  \label{thm:2} 
  The amanat protocol enforces secrecy (see Definition
  \ref{def:secrecy}) in all its executions unconditionally.
\end{theorem}

This means that \con is unable extract any piece of information on the
source \src which is not contained in \bin and \ocon in any case and
independently from any cryptographic assumptions. 

\begin{proof}
  During the execution of the protocol, \con receives the binary
  \bin, the output file \ocon, and the certificate \cert.
  The certificate $\cert=\sign(\kcpri, \langle \bin, \ocon
  \rangle,R_\round)$ is generated from \bin, \ocon, the key \kcpri,
  and the random seed $R_\round$. \con generates \kcpri itself and
  obtains access to \bin and to \ocon. Thus the only additional
  information communicated from \tbox to \pro stems from the random
  seed $R_\round$.
  But $R_\round$ has been fixed by \tbox before having access to \src
  and only depends on the iteration counter $\round$. Thus, \tbox
  cannot encode any information from \src into $R_\round$---such that
  the certificate depends on \bin and \ocon only.
\end{proof}

Note that the order of the random seeds $R_1,\dots,R_t$ must be
predetermined: Otherwise, \tbox could choose a random seed $R_i$
\emph{after} evaluating \src according to some conspirative scheme
\tbox and \con agreed upon.

%%% Local Variables:
%%% mode: latex
%%% TeX-master: "../cert-ip"
%%% ispell-local-dictionary: "american"  ***
%%% End:

\svnid{$Id: protocol-conformance.tex 241 2010-07-23 13:13:38Z schallha $}
\section{Protocol Conformance}
\label{sec:protocol-conformance}

We prove the conformance of our protocol using standard cryptographic
assumptions:
Following~\cite{goldreich04:_found_crypt}, we assume that the
public-key encryption is \emph{semantically secure} and that the
signature scheme is \emph{secure against adaptive chosen message
  attacks}, such as the RSA-based scheme proposed
in~\cite{cramer00:_signat_schem_based_strin_rsa_assum}.
Assuming these security properties, we obtain the conformance of our
protocol under all practically relevant conditions.

\begin{theorem}[Conformance]
  \label{thm:conformance} If the protocol terminates (in Step
  \textbf{C4} of the certification phase) with the customer \con
  accepting the certificate, then the protocol execution has been
  conformant in all but a negligible fraction of the cases.
  
  We assume that (a) the underlying encryption is semantically secure,
  (b) the signature scheme is secure against adaptively chosen message
  attacks, and (c) the supplier \pro runs in probabilistic polynomial
  time.
\end{theorem}

To prove this theorem, we recall in
Section~\ref{sec:security-properties} the stated security properties,
discuss in Section~\ref{sec:proof-overview} the proof structure, and
concretize the proof throughout Sections~\ref{sec:modeling-supplier}
to~\ref{sec:proof-theorem-ref}.

\subsection{Security Properties}
\label{sec:security-properties}

\emph{Semantic security} means that a probabilistic polynomial time
adversary cannot learn more from the ciphertext than from the length
of the plaintext alone.

\begin{definition}[Semantic Security~\cite{goldreich04:_found_crypt}]
  \label{def:semantic-security}
  A public-key encryption scheme (Definition~\ref{def:encryption}) is
  \emph{semantically secure} if for every probabilistic polynomial
  time algorithm $A$, there exist a probabilistic polynomial time
  algorithm $A'$ and an integer $N$, such that for every choice of
  $X_n$, $f$, $h$, $p$, $n\ge N$, and randomly chosen public key
  \kpub,
  $$
  \begin{array}{l}
    \Pr\left[A(1^n,\kpub,\kpub(X_n),1^{|X_n|},h(1^n,X_n))=f(1^n,X_n)\right]  \\
    < \Pr\left[A'(1^n,1^{|X_n|},h(1^n,X_n))=f(1^n,X_n)\right]  + 1/p(n)
  \end{array}
  $$
  holds. Therein, $X_n$ is a random variable of arbitrarily
  distributed plaintexts, $f$ and $h$ are functions in the security
  parameter $n$ and the plaintext $X_n$ which both yield a result of
  polynomial length, and $p$ is a polynomial.
\end{definition}

In this definition, we have two procedures $A$ and $A'$, where $A'$
receives the same information as $A$---except for the public key
$\kpub$ and the ciphertext $\kpub(X_n)$ which are dropped.

In many attacks on a public-key encryption scheme, the attacker does
not only receive some ciphertexts but also obtains some related
information. To model this situation, the function $h(1^n,X_n)$ is
used to provide both, $A$ and $A'$, with further information depending
on the plaintext $X_n$.

Then an encryption scheme is defined as semantically secure, if for
every $A$ there exists an $A'$ such that the probability that $A$ and
$A'$ differ in their results is at most negligible, i.e., both compute
the same function up to a negligible fraction of the cases.
Thereby, we say that a function $f:\mathbb{N}\rightarrow\mathbb{R}$ is
\emph{negligible}, iff there exits an integer $N$ for every polynomial
$p$ such that for all $n\ge N$, $f(n)<1/p(n)$ holds. When a
probabilistic experiment parameterized with $n$ has \emph{negligible
  success probability}, then we mean that this success probability as
function in the parameter $n$ is negligible. See
\cite{goldreich04:_found_crypt} for more details.

%% Finally, we point out that the functions $f$ and $h$ are only
%% restricted to produce a result of polynomial length.
%% %
%% In particular, they do not need to be computable at all.
%% 
%% \begin{remark}[Non-Uniformity in Semantic Security]
%%   \label{rem:arbitrary-advice}
%%   The functions $f$ and $h$ in Definition~\ref{def:semantic-security}
%%   are \emph{not required to be computable.} In particular, $f$ and $h$
%%   can \emph{use and provide arbitrary supplemental information.}
%% \end{remark}

An \emph{adaptive chosen message attack} is an attack against a
signature scheme, where the attacker is given a public key and access
to an oracle which can sign arbitrary messages with the corresponding
private key. 
The attacker generates a number of messages to be signed by the
oracle. The generated messages may depend on each other, on the public
key itself, and---since the attack is adaptive---on the signatures
previously returned by the oracle.
The attack procedure must then compute a message and a corresponding
signature which has \emph{not been signed before by the oracle.}

If every probabilistic polynomial time attacker is only successful in
forging a signature \emph{with a negligible probability,} then we say
that the signature scheme is \emph{secure against adaptive chosen
  message attacks.}

\begin{definition}[Adaptive Chosen Message
  Attack~\cite{cramer00:_signat_schem_based_strin_rsa_assum}] 
  \label{def:adaptive-chosen-message-attack}
  $\;\;$ Given a public key-pair $\left<\kpri,\kpub\right>$ for a signature
  scheme,
  a \emph{signing oracle} $S[\kpri]$ with private key \kpri is a
  function which takes a message $m$ and returns a signature
  $s=\sign(\kpri,m,R)$ for a uniformly and randomly chosen random seed
  $R$.
  A forging algorithm $F(\kpub)$ receives the public key \kpub and
  has access to the signing oracle $S[\kpri]$, where \kpri is the
  private key corresponding to \kpub.
    
  The algorithm $F$ is allowed to query $S[\kpri]$ for an arbitrary
  number of signatures. $F$ can \emph{adaptively choose the messages}
  to be signed, i.e., each newly chosen message can depend on the
  outcome of the previous queries.
  At the end of the computation, a \emph{successful attack} $F$ must
  output a message $m$ and a signature $s$ such that
  $\verifysignature(\kpub,m,s)$ succeeds, although $m$ has never been
  sent to and signed by $S[\kpri]$.

  A signature scheme (Definition~\ref{def:signatures}) is \emph{secure
    against adaptive chosen message attacks,} if every probabilistic
  polynomial time algorithm $F$ has only a negligible success
  probability.
\end{definition}

Assuming that the encryption scheme is semantically secure and that
the signature scheme is secure against adaptive chosen message
attacks, we start the conformance proof.

\subsection{Proof Overview}
\label{sec:proof-overview}

For the sake of contradiction, we assume that \pro is able to trick
\con into accepting a forged pair $\langle \bin,\ocon\rangle$ with a
not negligible success probability, i.e., \pro computes a certificate
\cert for a pair $\langle \bin,\ocon\rangle$ which has not been
computed and signed by \tbox but is nevertheless accepted by \con.
Starting from this assumption, we derive a contradiction in four
steps.

\paragraph*{1.~Modeling the Supplier
  (Section~\ref{sec:modeling-supplier})} As a first step in the proof,
we introduce in Fact~\ref{fact:modeling-the-supplier-in-c1} and
\ref{fact:modeling-the-supplier-in-c3} two procedures \progen and
\proval which model the computations of a honest supplier in Steps
\textbf{C1} and \textbf{C3} respectively.
Then each step of the supplier \pro corresponds to a single call to
one of these two procedures:
In each call, we provide the procedure with all information which has
been made available to \pro during the protocol execution so
far. After each call, the procedures \progen and \proval return all
information which is necessary to continue the protocol execution.
Since \pro can maintain an internal state between the individual
protocol steps, we also introduce a variable \state which is given to
both procedures \progen and \proval by reference, i.e., they can
access and update this variable \state.

\paragraph*{2.~A first Forging Procedure
  (Section~\ref{sec:first-forg-proc})} Since a malicious supplier is
assumed to exist, there must exist a corresponding pair of procedures
\mprogen and \mproval implementing the malicious behavior of this
supplier:
These procedures \mprogen and \mproval are only restricted to support
the same interface as \progen and \proval and to respect a
probabilistic polynomial time bound.
We use \mprogen and \mproval to build a first forging procedure
$\forge_1$:
This procedure simulates the possibly repeated certificate phase of
the Amanat protocol such that \mprogen and \mproval cannot distinguish
this simulation from a real protocol execution. 
Since they are unable to distinguish the simulation from a real
protocol execution, they will behave exactly the same way---and
consequently, in Proposition~\ref{lem:forge} we conclude that
$\forge_1$ produces the same result as a real protocol execution.
In the remainder of the proof, this procedure is transformed via an
intermediate procedure $\forge_2$ into the final attack algorithm
$\forge_3$.

\paragraph*{3.~Removing the Random Seeds and the Private Key
  (Section~\ref{sec:remov-encrpyt-rand})}
The procedure $\forge_1$ is not directly usable to forge certificates,
since it uses the private \kcpri which is supposed to be unknown to
$\forge_1$.
Hence, we want to use the signing oracle $S[\kcpri]$ instead; but when
we rely on $S[\kcpri]$ to generate certificates, we must also avoid
referring to the yet unused certificate random seeds
$R_{\round+1},\dots,R_t$ (since the choices of $S[\kcpri]$ for these
seeds are unpredictable).
Within $\forge_1$, $\mprogen$ and $\mproval$ both receive the session
key and the random seeds in an encrypted manner, i.e., they take
$\kmpub(\kcpri)$ and
$\krpub^{\round+1}(R_{\round+1}),\dots,\krpub^t(R_t)$ as arguments.
Because $\kcpri$ and $R_{\round+1},\dots,R_t$ are passed on in an
encrypted manner, we can apply semantic security to remove the
respective arguments
(Lemmata~\ref{lem:removing-encrypted-seeds-progen} and
\ref{lem:removing-encrypted-seeds-proval}).
In particular, we prove the existence of two procedures
$\overline{\mprogen}$ and $\overline{\mproval}$: They receive the same
arguments as their original counterparts---except for the encrypted
session key and the encrypted, yet unused random seeds---but return
the same output as \mprogen and \mproval, respectively.
Next, in Lemma~\ref{lem:polytime-substitutions} and
Corollary~\ref{lem:success-preservation}, we show that we can
substitute $\overline{\mprogen}$ and $\overline{\mproval}$ for
\mprogen and \mproval in any probabilistic polynomial time
procedure---again without changing the result of the procedure.
Lemmata~\ref{lem:removing-encrypted-seeds-progen}
to~\ref{lem:polytime-substitutions} hold in all but a negligible
fraction of the cases.

\paragraph*{4.~The Proof of Theorem~\ref{thm:conformance}
  (Section~\ref{sec:proof-theorem-ref})}
Lemma~\ref{lem:polytime-substitutions} applies to $\forge_1$ since it
is a probabilistic polynomial time procedure. Therefore, we replace in
$\forge_1$ \mprogen and \mproval with their counterparts
$\overline{\mprogen}$ and $\overline{\mproval}$ to obtain $\forge_2$.
Since $\forge_2$ uses the private key \kcpri for signing only and does
not make any use of the random seeds $R_{\round+1},\dots,R_t$, we can
replace all references to \kcpri with its signing oracle $S[\kcpri]$
to obtain $\forge_3$.
This procedure $\forge_3$ is the sought for adaptive chosen message
attack---contradicting the assumption that no such attack exists.

In the proof below, we assume that \emph{all procedures receive the
  security parameter $1^n$ implicitly} as their first parameter. We
also assume that their computations are polynomially bounded in $n$
and the length of their other inputs.

\subsection{Modeling the Supplier}
\label{sec:modeling-supplier}

To introduce the two procedures \progen and \proval which perform all
computations of the supplier \pro during the certification phase, we
observe that the supplier receives the following pieces of information
during the installation, session initialization, and \emph{possibly
  repeated} certification phase:
\begin{itemize}
\item The encrypted random seeds $\krpub^i(R_i)$ and the respective
  public keys $\krpub^i$ for $i=1,\dots,t$ are sent to \pro in Step
  \textbf{S6}.
\item The accumulated inputs \src, \bin, \opro, \ocon, \cert, and
  $\krpri^\round$ are sent to \pro at the end of Step \textbf{C2} of
  all previous certification phases.
\item All remaining messages of the initialization and session
  initialization phase comprise the following information: $\kmpub$
  (\textbf{I2}), $\kmpub(\kcpri)$ and \kcpub (\textbf{S1}), as well
  as $\cert_{\tool}$ and $\cert_{\comp}$ (\textbf{S2}).
\end{itemize}
We model the accumulated state of \pro with an additional variable
\state. This variable \state is given to \progen and \proval \emph{by
  reference,} i.e., both procedures and can access and update its
contents. Initially, \state contains all messages seen or generated by
\pro during the initialization and session initialization phase
($\kmpub$, \kcpub, $\cert_{\comp}$, and $\cert_{\tool}$). The
encrypted key $\kmpub(\kcpri)$ is handled explicitly and is therefore
not added to \state.
During a number of repeated certification phases, \pro can also
accumulate and store information on the received instances of \src,
\bin, \opro, \ocon, and \cert in \state. 

\begin{fact}[Modeling \pro for Step \textbf{C1}: \progen]
  \label{fact:modeling-the-supplier-in-c1}
  The computation of the supplier \pro in Step \textbf{C1} of the
  certification phase are modeled with a call
  $$
  \begin{array}{lll}
    \src=\progen & (\state, \kmpub(\kcpri), & \\
             & \krpub^1,\dots,\krpub^t, & \\
             & \krpub^1(R_1),\dots,\krpub^t(R_t), & \\
             & \krpri^1,\dots,\krpri^\round) & 
           \end{array}
  $$
  where \state is the state of \pro, $\kmpub(\kcpri)$ is the
  encrypted session key, \round is the protocol parameter
  being incremented with each iteration of the certification phase,
  and $t$ is the total number of precomputed random seeds.
\end{fact}

%% We deal with Step \textbf{C3} analogously:
%% %
%% Since the variable \round is incremented in Step \textbf{C2}, the
%% sequence $\krpri^1,\dots,\krpri^\round$ of private keys in the
%% invocation of \proval, described below, contains one more key than the
%% preceding invocation of \progen.

\begin{fact}[Modeling \pro for Step \textbf{C3}: \proval]
  \label{fact:modeling-the-supplier-in-c3}
  The computation of the supplier \pro in Step \textbf{C3} of the
  certification phase can be modeled with a call
  $$
  \begin{array}{lll}
    \result=\proval & (\state, \kmpub(\kcpri), \cert, & \\
             & \krpub^1,\dots,\krpub^t, & \\
             & \krpub^1(R_1),\dots,\krpub^t(R_t), & \\
             & \krpri^1,\dots,\krpri^\round) & 
           \end{array}
  $$
  where \state is the state of \pro, $\kmpub(\kcpri)$ is the
  encrypted session key, \cert is the certificate produced for the
  formerly generated \src, \round is the protocol parameter being
  incremented with each iteration of the certification phase, $t$ is
  the total number of precomputed random seeds, and \result indicates
  either
  \begin{itemize}
  \item to continue the protocol at Step \textbf{C4} (in this case
    \result contains a pair $\left<\bin,\ocon\right>$ and a
    corresponding but \emph{possibly forged} \cert to be sent to the
    \con),
  \item to abort the protocol and start again at Step \textbf{C1}, or 
  \item to refuse to work with \tbox since a secrecy violation has
    been detected.
  \end{itemize}
\end{fact}

In Fact~\ref{fact:modeling-the-supplier-in-c3}, \proval is only
receiving \cert while \pro receives in Step \textbf{C2} \bin, \opro,
and \ocon, as well. However, \proval can compute these pieces of
information itself with $\bin=\comp(\src)$ and
$\left<\opro,\ocon\right>=\tool(\src)$. To do so, \progen stores \src
in \state.

Note that we have to call \progen and \proval in an alternating manner
in order to ensure that these two procedures cannot distinguish a
simulation from a real protocol execution: This is necessary since
\state can be used to communicate information from \progen to \proval
and since they are called in an alternating manner in the original
protocol.

\subsection{A first Forging Procedure}
\label{sec:first-forg-proc}

We assume that there exists a malicious supplier \msup which produces
a pair $\langle \bin,\ocon\rangle$ and a certificate \cert within
probabilistic polynomial time at a \emph{not negligible probability}
such that
\begin{itemize}
\item \con accepts the pair and its certificate in Step \textbf{C4},
  but
\item \tbox did not produce the certificate \cert for $\langle
  \bin,\ocon\rangle$.
\end{itemize}
We model this malicious supplier with two procedures \mprogen and
\mproval, which run in probabilistic polynomial time and have the same
interface as \progen and \proval, as described in
Facts~\ref{fact:modeling-the-supplier-in-c1} and
\ref{fact:modeling-the-supplier-in-c3}.

Now we use the two malicious procedures to build a probabilistic
polynomial time forging algorithm $\forge_1$.
More specifically, $\forge_1$ embeds \mprogen and \mproval in a
simulated execution of the amanat protocol, consisting of a single
execution of the session initialization phase and a possibly repeated
execution of the certification phase.
Since \mprogen and \mproval cannot distinguish the simulation from the
real protocol execution, the success probability of $\forge_1$ is
identical to the success probability of the original malicious
supplier.
More specifically, the procedure $\forge_1$ takes as input a session
key-pair $\left<\kcpri,\kcpub\right>$ and---given the malicious
character of \mprogen and \mproval---produces with not negligible
probability a forged signature for this key-pair:
After computing with $\kcpri$ a number of signatures, $\forge_1$ uses
the accumulated information to \emph{forge} a signature, i.e., it
computes a signature without using $\kcpri$.
However, \mprogen and \mproval do require access to the encrypted
session key and random seeds---which is the problem addressed in the
following sections.

\begin{enumerate}
\item[$\mathbf{F1_1}$] {\em\bfseries Key and Random Seed Generation} \\
  Simulate the session initialization phase:
  \begin{itemize}
  \item Generate a sequence\\ $\langle
    \krpri^1,\krpub^1\rangle,\dots,\langle \krpri^t,\krpub^t\rangle$
    of key pairs.
  \item Generate a sequence $R_1,\dots,R_t$ of random seeds.
  \item Generate a master key pair $\left<\kmpri,\kmpub\right>$.
  \item Set $\round=0$.
  \item Initialize \state with $\kmpub$, \kcpub, $\cert_{\comp}$,
    and $\cert_{\tool}$.
  \end{itemize}

\item[$\mathbf{F2_1}$] {\em\bfseries Source Code Computation}  \\
  Simulate Step \textbf{C1} with a call
  $$
  \begin{array}{lll}
    \src=\mprogen & (\state, \kmpub(\kcpri), & \\
    & \krpub^1,\dots,\krpub^t, & \\
    & \krpub^1(R_1),\dots,\krpub^t(R_t), & \\
    & \krpri^1,\dots,\krpri^\round). & 
  \end{array}
  $$

\item[$\mathbf{F3_1}$] {\em\bfseries Source Code Verification} \\
  Simulate Step \textbf{C2} by computing 
  \begin{itemize}
  \item $\langle \opro, \ocon\rangle=\tool(\src)$ and
  \item $\bin=\comp(\src)$,
  \item incrementing \round by 1, and
  \item computing the certificate\\ $\cert=\sign(\kcpri,\langle
    \ocon,\bin\rangle,R_\round)$.
  \end{itemize}
  
\item[$\mathbf{F4_1}$] {\em\bfseries Secrecy Validation (Forge Certificate)}\\
  Simulate Step \textbf{C3}, i.e., make a call 
  $$
  \begin{array}{lll}
    \result=\mproval & (\state, \kmpub(\kcpri), \cert, & \\
    & \krpub^1,\dots,\krpub^t, & \\
    & \krpub^1(R_1),\dots,\krpub^t(R_t), & \\
    & \krpri^1,\dots,\krpri^\round). & 
  \end{array}
  $$
  Depending on \result, the execution proceeds with
  \begin{itemize}
  \item Step $\mathbf{F5_1}$, if \result indicates to continue the
    protocol (in this case \result contains $\langle
    \bin,\ocon\rangle$ and \cert, where \cert is forged with not
    negligible probability).
  \item Step $\mathbf{F2_1}$, if \result indicates to start over
    again.
  \item an \textbf{erroneous abort}, if \result indicates that a
    secrecy violation has been detected.
  \end{itemize}

\item[$\mathbf{F5_1}$] {\em\bfseries Output Result}\\
  Output the pair $\langle \bin,\ocon\rangle$ and \cert as indicated
  by \result.
\end{enumerate}

The procedure $\forge_1$ simulates the repeated execution of the
certification phase after a preceding initialization and session
initialization phase.
The procedure \mprogen and \mproval cannot distinguish between the
protocol execution and the simulation within $\forge_1$ and therefore,
$\forge_1$ produces a forged certificate in Step $\mathbf{F5_1}$ with
the same probability as the original supplier \msup in a protocol
execution. This leads to the following proposition:

\begin{proposition}[The Procedure $\forge_1$]
  \label{lem:forge}
  Let \mprogen and \mproval model a malicious supplier which is able
  to produce a non-conformant pair $\left<\bin,\ocon\right>$ with a
  valid (i.e., forged) certificate \cert with not negligible
  probability.
  Then, with non negligible probability, the procedure $\forge_1$
  outputs a forged certificate in Step $\mathbf{F5_1}$, i.e., one that has
  not been signed before in Step $\mathbf{F3_1}$.
\end{proposition}

However, both procedures \mprogen and \mproval are provided with the
encrypted session key $\kmpub(\kcpri)$ and the sequence
$\krpub^1(R_1),\dots,\krpub^t(R_t)$, i.e., \pro receives in advance
all random seeds which are used subsequently for certificates.
Since we want to use this procedure to mount an adaptive chosen
messages attack (Definition~\ref{def:adaptive-chosen-message-attack}),
we need to replace all direct references to \kcpri with queries to a
signing oracle $S[\kcpri]$.
But a signing oracle chooses its random seeds independently, and
therefore we have to cut down the sequence
$\krpub^1(R_1),\dots,\krpub^t(R_t)$ to
$\krpub^1(R_1),\dots,\krpub^\round(R_\round)$, i.e., no information on
the random seeds $R_{\round+1},\dots,R_t$ is allowed to be received by
\pro in advance.

\subsection{Removing the Random Seeds and the Private Key}
\label{sec:remov-encrpyt-rand}

As the random seeds
$\krpub^{\round+1}(R_{\round+1}),\dots,\krpub^t(R_t)$ and the private
key $\kmpub(\kcpri)$ are all encrypted, we can use the assumption that
the encryption scheme is semantically secure, as defined in
Definition~\ref{def:semantic-security}.
We first deal with \mprogen and then draw analogous conclusions for
\mproval.

\begin{lemma}[Removing Encrypted Seeds: $\overline{\mprogen}$]
  \label{lem:removing-encrypted-seeds-progen}
  For each probabilistic polynomial time procedure \mprogen which
  implements the signature of
  Fact~\ref{fact:modeling-the-supplier-in-c1},
  there exists another probabilistic polynomial time procedure
  $\overline{\mprogen}$ with the signature
  $$
  \begin{array}{lll}
    \src=\overline{\mprogen} & (\state, & \\
             & \krpub^1,\dots,\krpub^\round, & \\
             & \krpub^1(R_1),\dots,\krpub^\round(R_\round), & \\
             & \krpri^1,\dots,\krpri^\round) & 
           \end{array}
  $$
  such that for each polynomial $q$ and sufficiently large $n$
  $$
    \Pr\left[\overline{\mprogen}(\dots)=\mprogen(\dots)\right]>1-1/q(n) 
  $$
  holds.
\end{lemma}

\begin{proof}
  Fix a procedure \mprogen and a protocol parameter $t$ which is set
  to the maximum number of certification phases, i.e., the number of
  pre-committed random seeds.
  Then we start with the $t$-th random seed $R_t$ and remove the
  references to $\krpub^t$ and $\krpub^t(R_t)$. Next, we remove
  references to $\krpub^{t-1}$ and $\krpub^t(R_{t-1})$, and so forth,
  until we reach $\krpub^{\round}$ and $\krpub^\round(R_{\round})$.

  To apply Definition~\ref{def:semantic-security}, we define $h$ as
  constant function with
  $$
  \begin{array}{lll}
    h(R_t)=& (\state, \kmpub(\kcpri), & \\
    & \krpub^1,\dots,\krpub^{t-1}, & \\
    & \krpub^1(R_1),\dots,\krpub^{t-1}(R_{t-1}), & \\
    & \krpri^1,\dots,\krpri^{t-1}). & 
  \end{array}
  $$
  After permuting some arguments of \mprogen, we can rewrite a call to
  \mprogen as
  $$
  \mprogen(\krpub^t,\krpub^t(R_t), h(R_t))
  $$
  and set
  $$
  f(R_t)=\mprogen(\krpub^t,\krpub^t(R_t), h(R_t)).
  $$
  Please recall that we omit the implicit security parameter in our
  procedure headers, since then---by adding the implicit security
  parameter $1^n$ again---this setting matches precisely the
  prerequisites of Definition~\ref{def:semantic-security}.
  Thus, by the semantic security of the encryption scheme, there
  exists a corresponding procedure $\mprogen'(h(R_t))$ with
  $$
  \begin{array}{ll}
    \Pr\left[\mprogen(\krpub^t,\krpub^t(R_t), h(R_t))= f(R_t)\right] &\\[1mm]
    < \Pr\left[\mprogen'(h(R_t))= f(R_t)\right] +\frac{1}{p(n)} 
  \end{array}
  $$
  for every polynomial $p$ and sufficiently large $n$. Since $f$
  computes \mprogen, the first probability equals 1, yielding 
  $$
  \Pr[ \mprogen'(h(R_t))= \mprogen(\dots)]> 1-\frac{1}{p(n)}.
  $$
  Then, for an arbitrary polynomial $q$, we set
  $p(n)=(t-\round+1)q(n)$ and iterate the above process $(t-\round)$
  times,
  to obtain---for arbitrary $q$ and sufficiently large $n$---
  $$
  \begin{array}{ll}
    & \Pr[ \mprogen''(\dots)= \mprogen(\dots)] \\[2mm]
    > & \left(1-\frac{1}{p(n)}\right)^{t-\round} 
    =  \left(1-\frac{1}{(t-\round+1)q(n)}\right)^{t-\round} \\[2mm]
  \end{array}
  $$
  where we call $\mprogen''(\dots)$ with signature
  $$
  \begin{array}{lll}
     \mprogen'' & (\state, \kmpub(\kcpri), & \\
       & \krpub^1,\dots,\krpub^{\round}, & \\
       & \krpub^1(R_1),\dots,\krpub^{\round}(R_{\round}), & \\
       & \krpri^1,\dots,\krpri^\round). & 
  \end{array}
  $$
  It remains to remove $\kmpub(\kcpri)$ which can be done with one
  further application of the encryption scheme's semantic
  security. Then we arrive at
  $$
  \begin{array}{ll}
    & \Pr[ \overline{\mprogen}(\dots)= \mprogen(\dots)] \\[2mm]
    > & \left(1-\frac{1}{(t-\round+1)q(n)}\right)^{t-\round+1} \\[2mm]
    \ge & 1-\frac{1}{q(n)}
  \end{array}
  $$
  which is the Lemma statement.
\end{proof}

By substituting \mproval for \mprogen, we obtain in a completely
analogous manner the next lemma:

\begin{lemma}[Removing Encrypted Seeds: $\overline{\mproval}$]
  \label{lem:removing-encrypted-seeds-proval}
  For each procedure \mproval which runs in probabilistic polynomial
  time and implements the signature of
  Fact~\ref{fact:modeling-the-supplier-in-c3},
  there exists another probabilistic polynomial time procedure
  $\overline{\mproval}$ with the signature
  $$
  \begin{array}{lll}
    \result=\overline{\mproval} & (\state, \cert, & \\
             & \krpub^1,\dots,\krpub^\round, & \\
             & \krpub^1(R_1),\dots,\krpub^\round(R_\round), & \\
             & \krpri^1,\dots,\krpri^\round) & 
           \end{array}
  $$
  such that for each polynomial $q$ and sufficiently large $n$
  $$
    \Pr\left[\overline{\mproval}(\dots)=\mproval(\dots)\right]>1-1/q(n) 
  $$
  holds. 
\end{lemma}

Thus, starting from a malicious supplier which is modeled by two
probabilistic polynomial time procedures \mprogen and \mproval, we
obtain by application of
Lemmata~\ref{lem:removing-encrypted-seeds-progen} and
\ref{lem:removing-encrypted-seeds-proval} two other probabilistic
polynomial time procedures $\overline{\mprogen}$ and
$\overline{\mproval}$ which both
\begin{itemize}
\item do not receive any encrypted random seed---besides those which
  have been used already,
\item do not receive the encrypted key $\kmpub(\kcpri)$, and
\item return the same result as their original counterparts in all but
  a negligible fraction of the cases.
\end{itemize}
The same holds true for general polynomial time computations which
involve \mprogen and \mproval:

\begin{lemma}[Substituting $\overline{\mprogen}$ and
  $\overline{\mproval}$]
  \label{lem:polytime-substitutions}
  Let $A_1$ be a probabilistic polynomial time algorithm which invokes
  \mprogen and \mproval (as defined in
  Fact~\ref{fact:modeling-the-supplier-in-c1} and
  \ref{fact:modeling-the-supplier-in-c3}). Furthermore, let $A_2$ be
  the procedure obtained from $A_1$ by substituting
  $\overline{\mprogen}$ and $\overline{\mproval}$ (taken from
  Lemmata~\ref{lem:removing-encrypted-seeds-progen} and
  \ref{lem:removing-encrypted-seeds-proval}) for \mprogen and
  \mproval, respectively.
  Then $A_2$ and $A_1$ compute deviating results with a negligible
  probability only.
\end{lemma}

\begin{proof}
  Since $A_1$ and $A_2$ run within polynomial time, they can only
  invoke their respective subprocedures a polynomial number of times
  and therefore, their results deviate only if at least one of their
  polynomial many subprocedure invocations deviates.

  But if we repeat an experiment with negligible success probability a
  polynomial number of times, then the probability of observing at
  least one successful attempt is still negligible (see for example
  the introduction of~\cite{goldreich04:_found_crypt}).

  Thus, invoking $\overline{\mprogen}$ and $\overline{\mproval}$ a
  polynomial number of times will produce deviations only with
  negligible probability---and consequently, $A_1$ and $A_2$ produce
  the same result in all but a negligible fraction of the cases.
\end{proof}

In the proof of Theorem \ref{thm:conformance}, we use the following
consequence of Lemma~\ref{lem:polytime-substitutions}:
\begin{corollary}[Preserving a Not Negligible Success Prob.]
  \label{lem:success-preservation}
  Let the procedures $A_1$ and $A_2$ be given as described in
  Lemma~\ref{lem:polytime-substitutions}.
  Then $A_1$ computes a result successfully with a not negligible
  probability iff $A_2$ has a not negligible success probability for
  the same computation.
\end{corollary}

\begin{proof}
  We show equivalently that $A_1$ computes some result with a
  negligible probability iff $A_2$ has a negligible success
  probability for the same computation.

  For $A_2$ to be successful, one of two cases must arise: Either
  $A_1$ is successful (and the results of $A_1$ and $A_2$ coincide),
  or the results of $A_1$ and $A_2$ deviate (and $A_1$ is not
  successful).
  Thus, the success probability of $A_2$ is bounded by the probability
  that either $A_1$ is successful or that the results of $A_1$ and
  $A_2$ deviate.

  The success probability of $A_1$ is negligible, as well as the
  probability that the results of $A_1$ and $A_2$ deviate.
  Consequently, the probability that $A_2$ succeeds is negligible.
  By exchanging $A_1$ and $A_2$, the converse follows and the
  statement is proved for both directions.
\end{proof}

\subsection{Proof of Theorem \ref{thm:conformance}}
\label{sec:proof-theorem-ref}

In the final proof, we use $\forge_1$, as provided in
Section~\ref{sec:first-forg-proc} and apply
Lemma~\ref{lem:polytime-substitutions} to $\forge_1$ to obtain the
procedure $\forge_2$.
Then, $\forge_2$ has the following properties:
\begin{itemize}
\item The random seeds to be used in future rounds are not referenced
  in advance.
\item The result computed by $\forge_2$ deviates from $\forge_1$ only
  in a negligible fraction of the cases. By assumption, $\forge_1$
  produces a forged certificate with a \emph{not negligible
    probability,} and thus the same holds true for $\forge_2$.
\end{itemize}
It remains to replace all references to the private session key \kcpri
with queries to the signing oracle $S[\kcpri]$ to obtain a forging
procedure $\forge_3$ which matches
Definition~\ref{def:adaptive-chosen-message-attack} of an adaptive
chosen message attack:

\emph{Proof (of Theorem~\ref{thm:conformance}):}
We apply Lemma~\ref{lem:polytime-substitutions} to $\forge_1$ as
defined in Section~\ref{sec:first-forg-proc} to obtain $\forge_2$
which produces forged certificates with not negligible probability by
Corollary~\ref{lem:success-preservation}. Below, we show the
transformed procedure $\forge_2$, which uses \kcpri only to generate
signatures.

\begin{enumerate}
\item[$\mathbf{F1_2}$] {\em\bfseries Key and Random Seed Generation} \\
  Simulate the session initialization phase:
  \begin{itemize}
  \item Generate a sequence\\ $\langle
    \krpri^1,\krpub^1\rangle,\dots,\langle \krpri^t,\krpub^t\rangle$
    of key pairs.
  \item Generate a sequence $R_1,\dots,R_t$ of random seeds.
  \item Generate a master key pair $\left<\kmpri,\kmpub\right>$.
  \item Set $\round=0$.
  \item Initialize \state with $\kmpub$, \kcpub, $\cert_{\comp}$,
    and $\cert_{\tool}$.
  \end{itemize}

\item[$\mathbf{F2_2}$] {\em\bfseries Source Code Computation}  \\
  Simulate Step \textbf{C1} with a call
  $$
  \begin{array}{lll}
    \src=\overline{\mprogen} & (\state, & \\
    & \krpub^1,\dots,\krpub^\round, & \\
    & \krpub^1(R_1),\dots,\krpub^\round(R_\round), & \\
    & \krpri^1,\dots,\krpri^\round). & 
  \end{array}
  $$

\item[$\mathbf{F3_2}$] {\em\bfseries Source Code Verification} \\
  Simulate Step \textbf{C2} by computing 
  \begin{itemize}
  \item $\langle \opro, \ocon\rangle=\tool(\src)$ and
  \item $\bin=\comp(\src)$,
  \item incrementing \round by 1, and
  \item computing the certificate\\ $\cert=\sign(\kcpri,\langle
    \ocon,\bin\rangle,R_\round)$.
  \end{itemize}
  
\item[$\mathbf{F4_2}$] {\em\bfseries Secrecy Validation (Forge Certificate)}\\
  Simulate Step \textbf{C3}, i.e., make a call 
  $$
  \begin{array}{lll}
    \result=\overline{\mproval} & (\state, \cert, & \\
    & \krpub^1,\dots,\krpub^\round, & \\
    & \krpub^1(R_1),\dots,\krpub^\round(R_\round), & \\
    & \krpri^1,\dots,\krpri^\round). & 
  \end{array}
  $$
  Depending on \result, the execution proceeds with
  \begin{itemize}
  \item Step $\mathbf{F5_2}$, if \result indicates to continue.
  \item Step $\mathbf{F2_2}$, if \result indicates to start over again.
  \item an \textbf{erroneous abort}, if \result indicates that a
    secrecy violation has been detected.
  \end{itemize}

\item[$\mathbf{F5_2}$] {\em\bfseries Output Result}\\
  Output the pair $\langle \bin,\ocon\rangle$ and \cert as indicated
  by \result.
\end{enumerate}

Based on $\forge_2$, we build the attack procedure $\forge_3$ with two
further modifications such that they do not change the not negligible
success probability of the overall procedure.
\begin{itemize}
\item First, in Step $\mathbf{F1_2}$, we drop the
  precomputation of the random seeds $R_1,\dots,R_t$. 
\item Second, in Step $\mathbf{F3_2}$, we rely on the signing oracle
  $S[\kcpri]$ to generate the certificate \cert, and obtain with
  \extract the random seed that has been used by the oracle to compute
  the last certificate.
\end{itemize}
The success probability remains the same, since $\overline{\mprogen}$
and $\overline{\mproval}$ cannot detected the modifications, and since
the random seeds $R_1,\dots,R_t$ are generated in both cases according
to the uniform distribution.
Below, we show the updated Steps $\mathbf{F1_3}$ and $\mathbf{F3_3}$,
while the other ones remain the same.

\begin{enumerate}
\item[$\mathbf{F1_3}$] {\em\bfseries Key and Random Seed Generation} \\
  Simulate the session initialization phase:
  \begin{itemize}
  \item Generate a sequence\\ $\langle
    \krpri^1,\krpub^1\rangle,\dots,\langle \krpri^t,\krpub^t\rangle$
    of key pairs.
  \item Generate a master key pair $\left<\kmpri,\kmpub\right>$.
  \item Set $\round=0$.
  \item Initialize \state with $\kmpub$, \kcpub, $\cert_{\comp}$,
    and $\cert_{\tool}$.
  \end{itemize}

\item[$\mathbf{F3_3}$] {\em\bfseries Source Code Verification} \\
  Simulate Step \textbf{C2} by computing 
  \begin{itemize}
  \item $\langle \opro, \ocon\rangle=\tool(\src)$ and
  \item $\bin=\comp(\src)$,
  \item incrementing \round by 1,
  \item computing the certificate with the signing oracle
    $\cert=S[\kcpri](\langle \ocon,\bin\rangle)$, and 
  \item extract the random seed $R_\round=\extract(\cert)$ from the
    certificate \cert.
  \end{itemize}
\end{enumerate}

Since $\forge_2$ has a not negligible chance to forge certificates,
and since the output of $\forge_3$ is not changed by the last two
changes, $\forge_3$ has a not negligible chance to forge certificates
as well.
Moreover, $\forge_3$ accesses the private key \kcpri only in terms of
the signing oracle $S[\kcpri]$, and runs in probabilistic polynomial
time.

In other words, $\forge_3$ is a successful adaptive chosen message
attack, as defined in
Definition~\ref{def:adaptive-chosen-message-attack}. This is a
contradiction and concludes the proof.
\hfill$\blacksquare$

%%% Local Variables:
%%% mode: latex
%%% TeX-master: "../cert-ip"
%%% ispell-local-dictionary: "american"  ***
%%% End:

\svnid{$Id: conclusion.tex 241 2010-07-23 13:13:38Z schallha $}
\section{Conclusion}
\label{sec:conclusion}

IP boundaries impose an obstacle in the dissemination and application
of verification techniques. In the commonly considered verification
scenario, a relationship of mutual trust between the software author
and the verification engineer is presumed:
First, the verification engineer believes that the sources provided by
the software author have been indeed used to produce the final binary,
and second, the software author expects the verification engineer to
respect its IP rights on the provided sources.

But in an industrial context, violated IP rights and forged
verification verdicts entail enormous monetary damages and henceforth
a mutual trust relationship is insufficient as protection against such
misconduct.
We identified two security properties which are essential for any
security solution facilitating verification across IP boundaries:
First, \emph{conformance} requires that the verification verdict and
the delivered binary are produced from the same source, 
and second, \emph{secrecy} requires that the customer does not learn
anything about the sources which is not already directly encoded
within the binary and the verdict.

Taking this situation as starting point, we introduced the \emph{amanat
protocol} as a solution which satisfies both, conformance as well as
secrecy.

Subsequently, we proved the secrecy of the protocol in an intuitive
and cryptographically unconditional manner. This is important as to
provide a simple argument which asserts the well-protection of the
involved IP in a manner which is convincing to engineering and
management staff of a code supplying company.
In case of conformance, the proof required a much more technical
approach since abstract reasoning on the protocol (e.g., following the
Dolev-Yao style) is insufficient to establish conformance.

% \TODO{CS: there was not much future work\dots what to say here?}
%
We also envision wider applications of our protocol: First, we 
consider deep supply chains in the B2B setting where the final code
consumer wants to ensure conformance while facing a possibly
maliciously colluding group of chained suppliers. 
Second, we consider a B2C setting, i.e., for commercial-off-the-shelf
software. In this case, the customer party of the amanat protocol will
not be enacted by an end customer, but by a certification agency which
provides commercial verification services. 
A detailed exploration of these scenarios will be part of future work.

\begin{acks}
  We are thankful to Josh Berdine and Byron Cook for discussions on
  the device driver scenario and to Andreas Holzer and Stefan Kugele
  for comments on early drafts of the paper.
\end{acks}

%---------------------------------------------------------------------%
\bibliographystyle{acmtrans}
\bibliography{cert-ip}

\end{document}